\documentclass[useAMS,usenatbib]{mn2e}
\usepackage{graphicx}
\usepackage{amssymb}
\usepackage{color}

\newcommand{\nw}{}%{\textbf}
\newcommand{\z}{\zeta}

\title[Intrinsic conditions and structure of AGN jets]%
{Intrinsic Physical Conditions and Structure of Relativistic Jets in Active Galactic Nuclei}
\author[E.~E.~Nokhrina, V.~S.~Beskin, Y.~Y.~Kovalev, and A.~A.~Zheltoukhov]
{E.~E.~Nokhrina$^{1}$\thanks{E-mail: nokhrinaelena@gmail.com (EEN)}, V.~S.~Beskin$^{1,2}$,
Y.~Y.~Kovalev$^{2,3}$ and A.~A.~Zheltoukhov$^{1,2}$
\\
$^{1}$Moscow Institute of Physics and Technology, Institutsky per.~9, Dolgoprudny 141700, Russia\\
$^{2}$Lebedev Physical Institute, Russian Academy of Sciences, Leninsky prospekt 53, Moscow 119991, Russia\\
$^{3}$Max Planck Institute for Radio Astronomy, Auf dem H\"ugel 69, D-53121 Bonn, Germany
}

\begin{document}

\date{Accepted 4 December 2014; Received 24 November 2014; in original form 22 September 2014}

\pagerange{\pageref{firstpage}--\pageref{lastpage}} \pubyear{2014}

\maketitle

\label{firstpage}

\begin{abstract}
The analysis of the frequency dependence of the observed shift of
the cores of relativistic jets in active galactic nuclei (AGN)
allows us to evaluate the number density of the outflowing plasma
$n_\mathrm{e}$ and, hence, the multiplicity parameter $\lambda =
n_\mathrm{e}/n_\mathrm{GJ}$, where $n_\mathrm{GJ}$ is the Goldreich-Julian
number density. We have obtained \nw{the median value for $\lambda_\mathrm{med}=3\cdot 10^{13}$ 
and the median value for the Michel magnetization parameter $\sigma_\mathrm{M,med}=8$
from an analysis of 97 sources.} Since the
magnetization parameter can be interpreted as the maximum possible
Lorentz factor $\Gamma$ of the bulk motion which can be obtained
for relativistic magnetohydrodynamic (MHD) flow, this estimate is
in agreement with the observed superluminal motion of bright
features in AGN jets. Moreover, knowing these key parameters, one
can determine the transverse structure of the flow. We show that
the poloidal magnetic field and particle number density are much
larger in the center of the jet than near the jet boundary. 
\nw{The MHD model can also explain the typical observed level of jet acceleration.}
Finally, casual connectivity of strongly collimated jets is discussed.
\end{abstract}

\begin{keywords}
galaxies: active ---
galaxies: jets ---
quasars: general ---
radio continuum: galaxies ---
radiation mechanisms: non-thermal
\end{keywords}

\section{Introduction}
\label{s:intro}

%%%%%%%%%%%%%%%%%%%%%%%%%%%%%%%%%%%%%%%%%%%%%%%%%%%%%%%%%%%%%
Strongly collimated jets represent one of the most visible signs of the activity
of compact astrophysical sources. They are observed both in relativistic objects such as
active galactic nuclei (AGNs) and microquasars, and in young stars where the motion of
matter is definitely nonrelativistic. This implies that we are dealing with some universal
and extremely efficient mechanism of energy release.

At present the magnetohydrodynamic (MHD) model of activity of compact objects
is accepted by most astrophysicists~\citep{Mestel, MHD1}. At the heart of the
MHD approach lies the model of the unipolar inductor, i.e., a rotating source
of direct current. It is believed that the electromagnetic energy
flux --- the Poynting flux --- plays the main role in the energy transfer
from the `central engine' to active regions. The conditions for the existence
of such a `central engine' are satisfied in all the compact sources mentioned
above. Indeed, all compact sources are assumed to harbor a rapidly spinning
central body (black hole, neutron star, or young star) and some regular magnetic
field, which leads to the emergence of strong induction electric fields. The
electric fields, in turn, lead to the appearance of longitudinal electric currents
resulting in effective energy losses and particle acceleration.

The first studies of the electromagnetic model of compact sources (namely, radio
pulsars) were carried out as early as the end of the 1960s~\citep{GJ, Mich69}. It
was evidenced that there are objects in the Universe in which electrodynamical
processes can play the decisive role in the energy release. Then,~\citet{B76}
and~\citet{L76} independently suggested that the same mechanism can also operate
in active galactic nuclei, and for nearly 40 years this model has remained the
leading one.

Remember that within the MHD approach the total energy losses $P_{\rm jet}$ can be
easily evaluated as $P_{\rm jet} \sim IU$, where $I$ is the total electric current
flowing along the jet, and \mbox{$U \sim E R_{0}$} is the \nw{electro-motive 
force exerted on the black hole} on the 
scale $R_{0}$. If the central engine (black hole for AGNs) rotates withthe angular 
velocity $\Omega$ in the external magnetic field $B_{0}$, one can evaluate the 
electric field as \mbox{$E \sim (\Omega R_{0}/c)B_{0}$.} On the other hand, assuming 
that for AGNs the current density $j$  is fully determined by the relativistic outflow 
of the Goldreich-Julian charge density $\rho_{\rm GJ} = \Omega B_{0}/(2\pi c)$ (i.e., 
the minimum charge density required for the screening of the longitudinal electric 
field in the magnetosphere), one can write down $j \approx c \rho_{\rm GJ}$. It gives
\begin{equation}
I \sim \Omega B_{0}R_{0}^2.\label{Iev}
\label{IGJ}
\end{equation}
As for AGNs, one can set $R_{0} \approx r_{\rm g} = 2GM/c^2$ to obtain the
well-known evaluation~\citep{BZ-77}
\begin{equation}
P_{\rm jet} \sim \left(\frac{\Omega r_{\rm g}}{c}\right)^{2} B_{0}^{2} r_{\rm g}^{2} c.\label{Pjetev}
\label{BZ}
\end{equation}
In particular, comparing expressions (\ref{Iev}) and (\ref{Pjetev}), one can straightforwardly obtain
\begin{equation}
I \approx c^{1/2}P_{\rm jet}^{1/2}.
\label{i_0}
\end{equation}
For AGNs it corresponds to $10^{19}$--$10^{21}$~A. Certainly, the question
as to whether it is possible to consider a black hole immersed in an external
magnetic field as a unipolar inductor turned out to be also rather nontrivial
\citep{Punsly, Okamoto, Beskinbook}.

As a result, the MHD model was successfully used to describe a lot of processes in
active nuclei including the problem of the stability of jets~\citep{Benford, H&N,
A&C, IP94, B&K, Lyu09} and their synchrotron radiation~\citep{BK79, PIB, Lyu05}.
In particular, it was shown both analytically~\citep{Bog-95, HN-03, BN-09}
and numerically \citep{Komissarov-06, Tchekhovskoy-09, Porth-11, McKinney-12}
that for sufficiently small ambient pressure the dense core can
be formed. This is related both to advances in the theory which have at last formulated
sufficiently simple analytical relations~\citep{BZ-77, Beskin-10}, and to the breakthrough
in numerical simulations~\citep{Komissarov-06, Tchekhovskoy-09, Porth-11, McKinney-12}
which confirmed theoretical predictions.

Moreover, recently~\citet{Kronberg} demonstrated that in QSO 3C303 the jet does possess
large enough toroidal magnetic field, the apropriate longitudinal electric current along
the jet $I \approx 1.7\times 10^{19}$~A being as large as the electric current $I_{\rm GJ}$
(Eq.~\ref{IGJ}) which is necessary to support the Poynting energy flux. Besides, the lack of
$\gamma$-ray radiation as probed by the \textit{Fermi} Observatory for AGN
jets observed at small
enough viewing angles $\vartheta < 5^{\circ}$~\citep{Savetal-10} can be easily explained as well.
Indeed, as was found (see, e.g., Beskin 2010), within well-collimated magnetically
dominated MHD jets the Lorentz factors of the particle bulk motion can be evaluated as
\begin{equation}
\Gamma \approx r_{\perp}/R_{\rm L}, \label{gamma}
\end{equation}
where $r_{\perp}$ is the distance from the jet axis, and $R_{\rm
L} = c/\Omega$ is the light cylinder radius. Thus, the energy of
particles radiating in small enough angles $\vartheta$ with
respect to the jet axis is to be much smaller than that
corresponding to peripheral parts of a jet.

The most important MHD parameters describing relativistic flows (which was
originally introduced for radio pulsars) are the Michel magnetization parameter
$\sigma_{\rm M}$ and the multiplicity parameter $\lambda$. The first one determines the
maximum possible bulk Lorentz factor $\Gamma$ of the flow when all the energy transported
by the Poynting flux is transmitted to particles. The second one is the dimensionless
multiplicity parameter $\lambda = n_{\rm e}/n_{\rm GJ}$, which is defined as the
ratio of the number density $n_{\rm e}$ to the Goldrech-Julian (GJ) number density
$n_{\rm GJ} = \Omega B/2 \pi c e$. It is
important that these two parameters are connected by the simple relation~\citep{Beskin-10}
\begin{equation}
\sigma_{\rm M} \approx \frac{1}{\lambda}\left(\frac{P_{\rm jet}}{P_{\rm
A}}\right)^{1/2}.
\label{newsigma}
\end{equation}
Here $P_{\rm A} = m^{2}c^{5}/e^{2} \approx 10^{17}$~erg/s is the minimum energy losses
of the central engine which can accelerate particles to relativistic energies, and $P_{\rm jet}$
is the total energy losses of the compact object.

Unfortunately, up to now \nw{neither the magnetization nor the multiplicity parameters were actually
known} as the observations could not give us the direct information about the number density
and bulk energy of particles. The core shift method has been applied to obtain the concentration
$n_{\rm e}$, magnetic field $B$~\citep{Lob-98, Sull-09,Pushetal-12, Zdziarskietal-14}, and the jet composition \citep{Hir-05}
in AGN jets. However, evaluation of multiplicity and Michel magnetization parameters, which needs
to estimate the total jet power, has not been done.  From a theoretical point of view if the
inner parts of the accretiondisc are hot enough, then electron-positron pairs can be produced by
two-photon collisions,where photons with sufficient energy originate from the inner parts of the
accretion disk~\citep{BZ-77, Mosetal-11}. In this case \mbox{$\lambda \sim 10^{10}$--$10^{13}$,}
and Michel magnetization parameter $\sigma_{\rm M} \sim 10$--$10^{3}$. The second model takes
into account the appearance of the region where the GJ plasma density is equal to zero due to
general relativity effects that corresponds to the outer gap in the pulsar magnetosphere
\citep{BIP-92, HirOk-98}. This model gives $\lambda \sim 10^{2}$--$10^{3}$,  and
$\sigma_{\rm M} \sim 10^{10}$--$10^{13}$.

This large difference in the estimates for the magnetization parameter $\sigma_{\rm M}$
leads to two completely different pictures of the flow structure in jets. In particular,
it determines whether the flow is magnetically or particle dominated. The point is that
for ordinary jets $r_{\perp}/R_{\rm L} \sim 10^4$--$10^5$. As a result, using the
universal asymptotic solution $\Gamma \approx r_{\perp}/R_{\rm L}$ (\ref{gamma}), one can
obtain that the values $\sigma_{\rm M} \sim 10$--$10^{3}$ correspond to the saturation
regime when almost all the Poynting flux $W_{\rm em}$ is transmitted to the particle kinetic
energy flux $W_{\rm part}$. On the other hand, for $\sigma_{\rm M} \sim 10^{12}$ the jet
remains magnetically dominated ($W_{\rm part} \ll W_{\rm em}$). Thus, the determination
of Michel magnetization parameter $\sigma_{\rm M}$ is the key point in the analysis of
the internal structure of relativistic jets.

The paper is organized as follows. In section~\ref{s:multi} it is shown that VLBI observations
of synchrotron self-absorbtion in AGN jets allow us to evaluate the number density of the
outflowing plasma $n_{\rm e}$ and, hence, the multiplicity parameter $\lambda$. We discuss the
source sample and present the result for multiplicity and Michel magnetization mparameters in
section~\ref{s:calculation}. \nw{The values $\lambda \sim 10^{13}$ obtained from the analysis of
97 sources shows that for most jets the magnetization parameter $\sigma_{\rm M} \lesssim 30$}.
Since the magnetization parameter is the maximum possible value of the Lorentz factor of the
relativistic bulk flow, this estimate is consistent with observed superluminal motion.In
section~\ref{s:struct} it is shown that for physical parameters determined above, the
poloidal magnetic field and particle number density are much larger in the center of the
jet than near its boundary. Finally, in section~\ref{s:connect} the casual connectivity
of strongly collimated supersonic jets is discussed. Throughout the paper, we use the
$\Lambda$CDM cosmological model with $H_0=71$~km~s$^{-1}$~Mpc$^{-1}$, $\Omega_m=0.27$, and
$\Omega_\Lambda=0.73$ \citep{Komatsu09}.

\section{The method} 
\label{s:multi}

\subsection{General relations}
\label{ss:Gr}

To determine the multiplicity parameter $\lambda$ and Michel magnetization parameter
$\sigma_{\rm M}$ one can use the dependence on the visible position of the core of the
jet from the observation frequency~\citep{Gou-79, BK79, Mar-83, Lob-98, Hir-05, KLPZ-08,
Sull-09, Sok-11, Pushetal-12}. This effect is associated with the absorption of the
synchrotron photon gas by relativistic electrons (positrons) in a jet.

Typically, the parsec-scale radio morphology of a bright AGN manifests a one-sided jet
structure due to Doppler boosting that enhances the emission of the approaching jet. The
apparent base of the jet is commonly called the ``core'', and it is often the brightest
and most compact feature in VLBI images of AGN. The VLBI core is thought to represent
the jet region where the optical depth is equal to unity.

We will employ the following model to connect the physical parameters at the jet launching
region with the observable core-shift. There is a magnetohydrodynamic relativistic outflow
of non-emitting plasma moving with bulk Lorentz factor $\Gamma$ and concentration $n_{\rm e}$
in the observer rest frame. On the latter we superimpose the flow of emitting particles with
distribution ${\rm d}n_{{\rm syn}*} = k_{{\rm e}*}\gamma_{*}^{-1+2\alpha}{\rm d}\gamma_{*}$,
$\gamma_{*}\in[\gamma_{\rm min*};\;\gamma_{\rm max*}]$. Here $n_{\rm syn*}$ is concentration
of emitting plasma, $k_{\rm e*}$ is concentration aplitude, and $\gamma_*$ is the emitting
particles' Lorentz factor. All the parameters with subscript `*' are taken in the non-emitting
plasma rest frame, i.e., in the frame which locally moves with the bulk Lorentz factor $\Gamma$.

We suppose that the emitting particles radiate synchrotron photons in the jet's magnetic field,
and these photons scatter off the same electrons, which lead to the photon absorption
\citep{Gou-79, Lob-98, Hir-05}. The corresponding turn-over frequency $\nu_\mathrm{m*}$, the
frequency at which the flux density $S_{\nu}$ has a maximum, can be evaluated using
expressions from \citet{Gou-79} as
\begin{equation}
\nu_\mathrm{m*}^{(5-2\alpha)}=\frac{c_{\alpha}^{2}(1-2\alpha)}{5(5-2\alpha)}\frac{e^{4}}{m^{2}c^{2}}\left(\frac{e}{2\pi
mc}\right)^{3-2\alpha}R_{*}^{2}B_*^{3-2\alpha}k_{\rm e*}^{2}.
\end{equation}
\nw{The function $c_{\alpha}\,(\cdot)$ is a composition of gamma-functions defined by \citet{Gou-79}}, and
for $\alpha=-1/2$ we have $c_{\alpha}\,(2)=1.396$. Constants $e$, $m$, and $c$ are the electron charge,
electron mass, and the speed of light correspondingly. Finally, $B_*$ is the magnitude of
disordered magnetic field in an emitting region with a characteristic size $R_{*}$ along the
line of sight.

Although we assume that the toroidal magnetic field dominates in the jet, an assumption of disordered magnetic
field in our opinion can be retained, because, for an optically thin jet, the photon meets both
directions of field. Thus, the mean magnetic field along the photon path is almost zero, which mimics
the behaviour of a disordered field. As a result, the parameters in the observer rest frame and
plasma rest frame are connected by the following equations:
\begin{equation}
\frac{\nu_{\rm m*}}{\nu_{\rm m}}=\frac{1+z}{\delta},
\end{equation}
\begin{equation}
R_{*}=\frac{2\,r_*\chi_*}{\sin{\varphi_*}}=\frac{2\,r\chi}{\delta\sin{\varphi}},
\end{equation}
\begin{equation}
B_{*}=\sqrt{B^2-E^2}\approx\frac{B_{\varphi}}{\Gamma}\approx\frac{B}{\Gamma},
\end{equation}
\begin{equation}
k_{\rm e*}=\frac{k_{\rm e}}{\Gamma},
\end{equation}
where $z$ is the red-shift, 
\begin{equation}
\delta=\frac{1}{\Gamma\left(1-\beta\cos\varphi\right)}\label{doppler}
\end{equation}
is the Doppler factor, $\chi$
is the jet half-opening angle, and $\varphi$ is a viewing angle.

Further, the number density of emitting electrons $n_{\rm syn}$ is connected with the amplitude
$k_{\rm e}$ as
\begin{equation}
k_{\rm e}=n_{\rm syn}\frac{2\alpha}{\gamma_{\rm max}^{2\alpha}-\gamma_{\rm min}^{2\alpha}},
\end{equation}
where $\gamma=\gamma_*\Gamma$. For $\alpha=-1/2$ we get
\begin{equation}
k_{\rm e}\approx n_{\rm syn}\gamma_{\rm min}.
\end{equation}
We also put $n_{\rm syn}=\xi n_{\rm e}$. Here $\xi$ is a ratio of the number density of emitting particles
to the MHD flow number density. The portion of particles effectively accelerated by the internal
shocks was found by \citet{SSA-13} to be about 1\%, so we take $\xi\approx 0.01$.

Finally, we assume~\citep{Lob-98, Hir-05} the following power law behavior for the magnetic field
and particle density dependence on distance:
\begin{equation}
B(r)=B_{1}\left(\frac{r}{r_{1}}\right)^{-1},
\end{equation}
\begin{equation}
n_{\rm e}(r)=n_1\left(\frac{r}{r_1}\right)^{-2},
\end{equation}
where $B_{1}$ is the magnetic field and $n_1$ is the number density at $r_{1}=1$ pc respectively.
For these scalings of particle density and magnetic field with the distance the turn-over frequency
$\nu_{\rm m}$ as a function of $r$ does not depend on $\alpha$ and can be written as
\begin{equation}
\nu_{\rm m}\propto r^{-1}.
\end{equation}
This scaling has been confirmed by \citet{Sok-11} in measurements of core-shifts for 20 AGNs made
for 9 frequencies each. Using these dependencies of magnetic field and particle number density of
distance $r$, we obtain in the observer rest frame
\begin{equation}
\begin{array}{l}
\displaystyle\left(\nu_{\rm m}\frac{1+z}{\delta}\frac{r}{r_1}\right)^{5-2\alpha}
=C\left(\frac{2e^2}{mc}\right)^2
\left(\frac{e}{2\pi mc}\right)^{3-2\alpha}\times \\
\displaystyle \times\left(\frac{r_1\chi}{\delta\sin\varphi}\,\xi\gamma_{\rm min}\right)^2
\Gamma^{-5+2\alpha} B_1^{3-2\alpha}n_1^2,
\end{array}
\end{equation}
where $C=c_{\alpha}^2(1-2\alpha)/5(5-2\alpha)$.

On the other hand, the values $B_1$ and $n_1$ can be related through introducing the flow
magnetization parameter $\sigma$ --- the ratio of Poynting vector to particle kinetic
energy flux at a given distance along the flow (see Appendix~\ref{a:mp}). Let us define
the magnetization $\sigma_{\xi}$ as a ratio of Poynting vector to the total kinetic energy
flux of emitting and non-emitting particles:
\begin{equation}
\sigma_{\xi}=\frac{\left|\vec{S}\right|}{\left|\vec{K}+\vec{K}_{\rm syn}\right|}.
\end{equation}
Here the kinetic energy flux of emitting electrons is
\begin{equation}
\left|\vec{K}_{\rm syn}\right|=\int_{\gamma_{\rm min}}^{\gamma_{\rm max}}
(\gamma m c^2)(|\vec{v}_{\rm p}|){\rm d}n_{\rm syn}=
k_{\rm e}mc^3F(\gamma_{\rm min},\gamma_{\rm max}),
\end{equation}
and function $F\,(\cdot)$ for $\alpha=-1/2$ is defined by the following expression:
\begin{equation}
\begin{array}{l}
\displaystyle F(\gamma_{\rm min},\gamma_{\rm max})=\left({\rm ch}^{-1}\gamma_{\rm max}-{\rm ch}^{-1}\gamma_{\rm min}\right)-\\ \ \\
\displaystyle-\left(\frac{\sqrt{\gamma^2_{\rm max}-1}}{\gamma_{\rm max}}-\frac{\sqrt{\gamma^2_{\rm min}-1}}{\gamma_{\rm min}}\right)\approx \ln\left(2\gamma_{\rm max}\right)-1.
\end{array}
\end{equation}
Estimating now the Poynting vector as
\begin{equation}
\left|\vec{S}\right|\approx\frac{cB^2_{\varphi}}{4\pi},
\end{equation}
and particle kinetic energy fluxes as
\begin{equation}
\left|\vec{K}+\vec{K}_{\rm syn}\right|\approx mc^3n_{\rm e}\left(\Gamma+\xi F\gamma_{\rm min}\right),
\end{equation}
we obtain the following relationship between magnetic field and particle number density:
\begin{equation}
B^2=\sigma_{\xi} 4\pi mc^2 n_{\rm e} \Gamma.\label{equi}
\end{equation}
In what follows, we neglect the term $\xi F\gamma_{\rm min}$ in comparison with $\Gamma$.
Further on we omit the index $\xi$. Using (\ref{equi}), we get
\begin{equation}
\begin{array}{l}
\displaystyle\left(\nu_{\rm m}\frac{1+z}{\delta}\frac{r}{r_1}\right)^{5-2\alpha}
=C\left(\frac{2e^2}{mc}\right)^2\left(\frac{e}{2\pi mc}\right)^{3-2\alpha}\times \\ \ \\
\displaystyle \times\left(4\pi mc^2\right)^{1.5-\alpha}\left(\frac{r_1\chi}{\delta\sin\varphi}
\,\xi\gamma_{\rm min}\right)^2\Gamma^{-5+2\alpha}\times \\ \ \\
\displaystyle\times \left(\sigma_1\Gamma_1\right)^{1.5-\alpha} n_1^{3.5-\alpha}.\label{FA1}
\end{array}
\end{equation}

As to the number density $n_1$, it can be defined through the multiplicity parameter $\lambda$
and total jet energy losses $P_{\rm jet}$ as (see Appendix~\ref{a:lambda} for more detail)
\begin{equation}
n_1=\frac{\lambda}{2\pi \left(r_1\chi\right)^2}\frac{mc^2}{e^2}\sqrt{\frac{P_{\rm jet}}{P_{\rm A}}}.\label{nWtot}
\end{equation}

\subsection{The saturation regime}

To determine the intrinsic parameters of relativistic jets, let us consider two cases for the
different magnetization at 1~pc. In what follows we assume that the flow at its base is highly,
or at least mildly, magnetized, i.e. $\sigma_{\rm M}\gg 1$.

First, we assume that up to the distance $r=1$~pc the plasma has been effectivily accelerated
so that the Poynting flux is smaller in comparison with the particle kinetic energy flux, i.e.,
$\sigma_1 \lesssim 1$.  In other words, the acceleration reached the saturation regime~\citep{Beskin-10}.
Combining now (\ref{gamma}) and (\ref{sigma}), it is easy to obtain that this case corresponds
to $\sigma_{\rm M} < 10^{5}$. Accordingly, the bulk Lorentz factor at $r=1$~pc can be evaluated as
\begin{equation}
\Gamma_1 \approx \sigma_{\rm M}.
\end{equation}
In this case Eqn.~(\ref{FA1}) can be rewritten as
\begin{equation}
\left(\nu_{\rm m}\frac{1+z}{\delta}\frac{r\chi}{c}\right)^{5-2\alpha}
=C \left(\frac{2^{-1.5+\alpha}}{\pi^{5-2\alpha}}\right)
\left(\frac{\xi\gamma_{\rm min}}{\delta\sin\varphi}\right)^2\lambda^{7-2\alpha}.
\end{equation}
Using now the relationship between the angular distance $\theta_{\rm d}$ and the distance from
the jet base $r$
\begin{equation}
r\sin\varphi=\theta_{\rm d}\frac{D_{\rm L}}{(1+z)^2},
\end{equation}
where $D_{\rm L}$ is the luminosity distance, we obtain
\begin{equation}
\begin{array}{l}
\displaystyle\left(\frac{\theta_{\rm d}}{\rm mas}\right)
=3.4\cdot 10^{-19}\left(\frac{D_{\rm L}}{\rm Gpc}\right)^{-1}
\left(\frac{\nu_{\rm m}}{\rm GHz}\right)^{-1}
{\delta(1+z)}\times\\ \ \\
\displaystyle\times\frac{\sin\varphi}{\chi}
\left(\frac{\xi\gamma_{\rm min}}{\delta\sin\varphi}\right)^{2/(5-2\alpha)}\lambda^{(7-2\alpha)/(5-2\alpha)}.
\end{array}
\end{equation}
This expression can be rewritten as a following relationship between the core position and the
observation frequency:
\begin{equation}
\left(\frac{\theta_{\rm d}}{\rm mas}\right)=\left(\frac{\eta}{\rm mas\cdot GHz}\right)
\left(\frac{\nu_{\rm m}}{\rm GHz}\right)^{-1}.
\end{equation}
Having the measured core-shift $\Delta r_{\rm mas}$ in milliarcseconds for two frequencies
$\nu_{\rm m,1}$ and $\nu_{\rm m,2}$, we obtain for $\alpha=-1/2$:
\begin{equation}
\begin{array}{l}
\displaystyle\lambda=7.3\cdot 10^{13}\left(\frac{\eta}{\rm mas\cdot GHz}\right)^{3/4}
\left(\frac{D_{\rm L}}{\rm Gpc}\right)^{3/4}\times\\ \ \\
\displaystyle\times\left(\frac{\chi}{1+z}\right)^{3/4}\frac{1}{(\delta\sin\varphi)^{1/2}}
\frac{1}{(\xi\gamma_{\rm min})^{1/4}}=
\\ \ \\
\displaystyle =2.3\cdot 10^{14}\left(\frac{\eta}{\rm pc\cdot GHz}\right)^{3/4}
\left(\frac{D_{\rm L}}{\rm Gpc}\right)^{3/4}\times\\ \ \\
\displaystyle\times\left(\frac{\chi}{1+z}\right)^{3/4}\frac{1}{(\delta\sin\varphi)^{1/2}}
\frac{1}{(\xi\gamma_{\rm min})^{1/4}}.
\end{array}\label{Fin1}
\end{equation}
Accordingly, using (\ref{newsigma}), we obtain
\begin{equation}
\begin{array}{l}
\displaystyle\sigma_{\rm M}=1.4\cdot\left[\left(\frac{\eta}{\rm mas\cdot GHz}\right)
\left(\frac{D_{\rm L}}{\rm Gpc}\right)\frac{\chi}{1+z}\right]^{-3/4}\times
\\ \ \\
\displaystyle\times\sqrt{\delta\sin\varphi}
\left(\xi\gamma_{\rm min}\right)^{1/4}\sqrt{\frac{P_{\rm jet}}{10^{45}{\rm erg\cdot s^{-1}}}}=
\\ \ \\
\displaystyle=0.44\cdot\left[\left(\frac{\eta}{\rm pc\cdot GHz}\right)
\left(\frac{D_{\rm L}}{\rm Gpc}\right)\frac{\chi}{1+z}\right]^{-3/4}\times
\\ \ \\
\displaystyle\times\sqrt{\delta\sin\varphi}
\left(\xi\gamma_{\rm min}\right)^{1/4}\sqrt{\frac{P_{\rm jet}}{10^{45}{\rm erg\cdot s^{-1}}}}.
\end{array}
\end{equation}
As we see, this value is in agreement with our assumption $\sigma_{\rm M} < 10^{5}$.

\subsection{Highly magnetized outflow}

Let us now assume that the flow is still highly magnetized at a distance of the observale core.
This implies that the Michel magnetization parameter $\sigma_{\rm M} > 10^5$. Using now relation
(\ref{sigma}), one can obtain
\begin{equation}
\sigma_1\Gamma_1\approx\sigma_{\rm M}.
\end{equation}
On the other hand, Eqn. (\ref{FA1}) can be rewritten as
\begin{equation}
\begin{array}{l}
\displaystyle\lambda=2.5\cdot 10^{11}\left[\left(\frac{\eta}{\rm mas\cdot GHz}\right)
\left(\frac{D_{\rm L}}{\rm Gpc}\right)\left(\frac{\chi}{1+z}\right)\right]^3\times \\ \ \\
\displaystyle\times
\Gamma^{3/2}\frac{1}{(\delta\sin\varphi)^2}\frac{1}{\xi\gamma_{\rm min}}
\left(\frac{P_{\rm jet}}{10^{45} \rm erg\cdot s^{-1}}\right)^{-3/2}.
\end{array}\label{Fin2}
\end{equation}
This gives the following expression for the Michel magnetization parameter
\begin{equation}
\begin{array}{l}
\displaystyle\sigma_{\rm M}=4\cdot 10^{2}
\left[\left(\frac{\eta}{\rm mas\cdot GHz}\right)
\left(\frac{D_{\rm L}}{\rm Gpc}\right)\left(\frac{\chi}{1+z}\right)\right]^{-3}\times \\ \ \\
\displaystyle\times
\Gamma^{-3/2}(\delta\sin\varphi)^2\xi\gamma_{\rm min}
\left(\frac{P_{\rm jet}}{10^{45} \rm erg\cdot s^{-1}}\right)^{2}.
\end{array}
\label{Fin3}
\end{equation}
As we see, these values are in contradiction with our assumption $\sigma_{\rm M} > 10^5$.
Thus, one can conclude that it is the saturation limit that corresponds to parsec-scale
relativistic jets under consideration.

\section{The statistics for multiplicity parameter}
\label{s:calculation}

Several methods can be applied to measure the apparent shift of the core position
as discussed by \cite{KLPZ-08}. As a result, a magnitude of the shift, designated by $\eta$,
can be measured and presented in units [\mbox{mas $\cdot$ GHz}] or [\mbox{pc $\cdot$ GHz}].
Knowing this quantity, one can use the expressions (\ref{Fin1})--(\ref{Fin3}) to estimate the
multiplicity and magnetezation parameters.

\subsection{The sample of objects}
\label{s:sample}

In our analysis we use the results of two surveys of the apparent core shift in AGN jets:
\citet{Sok-11} show results for 20 objects obtained from nine frequencies between 1.4 and
15.3~GHz (S-sample) and \citet{Pushetal-12} have results
for 163 AGN from four frequencies covering 8.1-15.3 GHz (P-sample). Of these we use only
those sources for which the apparent opening angle is known from \citet{Pushetal-09}.
As a result, 97 sources are left from the P-sample and 5 from the S-sample. Although all of S-sample
sources are in P-sample, we have included them as an independent measurment of core shift.
Moreover, for the objects 0215+015 and 1219+285 the two measurements of core-shift has been
made for two different epochs, and we included them too. This leaves us with 97 sources and
104 measurements of core shift.

The distance to the objects is determined from the redshift and accepted cosmology
model. For a Doppler factor we use the estimate $\delta\approx\beta_{\rm app}$, where
measured apparent velocity $\beta_{\rm app}$ is a ratio of apparent speed of a bright
feature in a jet to the speed of light. 
We believe this to be a good estimate because
\citet{Cohetal-07} have showed using Monte-Carlo simulations that the probability density
$p\left(\delta|\,\beta_{\rm app}\right)$ to observe a Doppler factor for a given apparent
velocity is peaked around unity. 
This is done under an assumption that the measured $\beta_{\rm app}$ does represent the underlying jet flow.
The redshifts $z$ and the apparent velocities $\beta_{\rm app}$ are taken from \citet{Listeretal-13}.

The value of observation angle $\varphi$ we obtain from the set of equations for Doppler
factor $\delta$ and apparent velocity
\begin{equation}
\beta_{\rm app}=\frac{\beta\sin\varphi}{1-\beta\cos\varphi}.\label{betaapp}
\end{equation}
Taking $\delta=\beta_{\rm app}$, we obtain from (\ref{doppler}) and (\ref{betaapp}) for the observation angle the relation
\begin{equation}
\varphi={\rm atan}\left(\frac{2\beta_{\rm app}}{2\beta_{\rm app}^2-1}\right).
\end{equation}
The half-opening angle related to the observed opening angle $\chi_{\rm app}$ as
\begin{equation}
\chi=\chi_{\rm app}\sin\varphi/2.\label{chi}
\end{equation}
We use the values for $\chi_{\rm app}$ derived by \citet{Pushetal-09} with typical errors of $1\fdg5$.
We also have chosen parameter $\gamma_{\rm min}=1$.

We evaluate the total jet power $P_{\rm jet}$ through the relationship \citep{Cavagnolo-10} between the luminosities
of jets in radio band and mechanical jet power, needed to form the cavities in surrounding gas.
The power law, found by \citet{Cavagnolo-10} for a range of frequencies 200 -- 400 MHz is
\begin{equation}
\left(\frac{P_{\rm jet}}{10^{43}\,\rm erg\cdot s^{-1}}\right)=3.5
\left(\frac{P_{200-400}}{10^{40}\,\rm erg\cdot s^{-1}}\right)^{0.64}.
\end{equation}
In order to find flux density measurements at the 92~cm band for each source we use the CATS database \citep{Verkh-97} which 
accumulates measurements at different epoches and from the different catalogues. The data which
we use in this paper were originally reported by \citet{WISH, TXS, SRCGh, Ku79r, Ku81r, 87GBM, LFVAR, WENSS}
with a typical flux density accuracy of about 10\%.

The typical error for core shift measurements in \citet{Pushetal-12} and
\citet{Sok-11} is $0.05$~mas. There are 23 objects in our sample that have the core shift
values less than $0.05$~mas. For them we have replaced the core shift values
by $0.05$~mas for our calculations for convenience of the $\lambda$ and $\sigma_\mathrm{M}$ analysis.

\subsection{Results and discussion}
\label{s:result}

\begin{figure}
\includegraphics[scale=0.90]{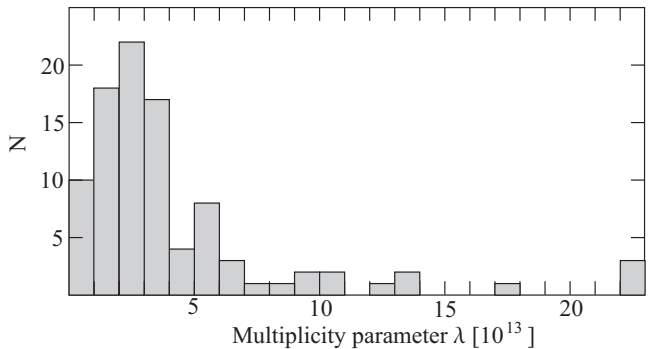}
\caption{
\nw{Distributions of the multiplicity parameter $\lambda$ for the sample of 97 sources. 
Two objects with $\lambda=2.8\cdot 10^{14}$ and $3.6\cdot 10^{14}$ lie out of the shown range of values.}
\label{hist1}
}
\end{figure}

\begin{figure}
\includegraphics[scale=0.90]{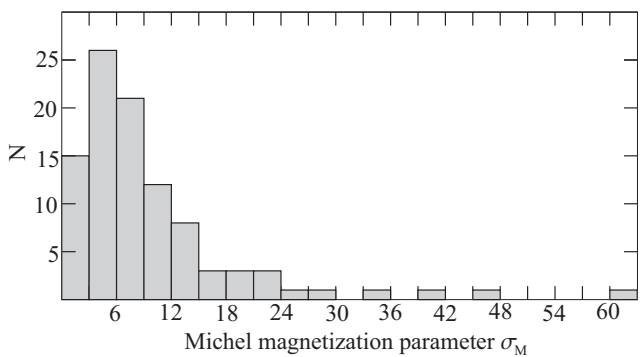}
\caption{
\nw{Distributions of the Michel magnetization parameter $\sigma_{\rm M}$ for the sample of 97 sources.}
\label{hist2}
}
\end{figure}

Using the formula (\ref{Fin1}), we obtain the following result for the equipartition regime.
The obtained values for the multiplication parameter $\lambda$ and magnetization parameter
$\sigma_\mathrm{M}$ are presented in Table~\ref{table}. 
Their distributions are shown in Fig.~\ref{hist1} and Fig.~\ref{hist2}, respectively.
In cases when more than one estimate is determined per source (e.g., for 0215+015), an average value is used in the histograms.
\nw{The resultant median value
for the multiplicity parameter $\lambda_{\rm med}=3\cdot 10^{13}$, and median value for
magnetization parameter $\sigma_{\rm M,\,med}=8$.  The multiplicity parameter for our
sample lies in the interval $(3\cdot 10^{12};\;4\cdot 10^{14})$, and Michel magnetization
parameter $\sigma_{\rm M}$ lies correspondingly in the $(0.4;\; 61)$ interval.}

The Doppler factor of a flow can be also obtained through the variability method by measuring
the amplitude and duration of a flare \citep{Hovetal-09}. Making an
assumption that the latter corresponds to the time needed for light to cross the emitting
region, and assuming the intrinsic brightness temperature is known (from the equipartition
argument), one can derive the beaming Doppler factor. 
We have used the variability Doppler factors obtained by \citet{Hovetal-09} for 50 objects with measured core shifts \citep{Pushetal-12}
instead of our original assumption for Doppler factor $\delta=\beta\mathrm{app}$ and have found 
that our estimates for $\lambda$ and $\sigma_\mathrm{M}$ stay the same within a factor of 2.

We estimate the total typical accuracy of $\lambda$ and $\sigma_\mathrm{M}$ values
in Table~\ref{table}
to be of a factor of a few. It is mostly due to the assumptions and simplifications
introduced and, to a less of an extent, due to accuracy of observational parameters of the jets.
We note that while an estimate for every source is not highly accurate, 
the distributions in Figs.~\ref{hist1},~\ref{hist2} should represent the sample properties well.

\nw{There are 3 objects in our sample that have Michel magnetization parameter $\sigma_{\rm M}<1$,
which means that the flow is not magnetically dominated at its base. And we have overall
9 sources with $\sigma_{\rm M}<2$, which is in contradiction with our assumption of at
least a mildly magnetized flow. This a small fraction (9\%) of all 97 sources, so we feel
that for the majority sources there is no contradiction of our assumptions and the resultant value
for Michel magnetization parameter.} % We believe that the small value of $\sigma_{\rm M}$ for
%these 15 sources may be attributed to our assumption $\delta\approx\beta_{\rm app}$, which
%is equivalent to $\sin\varphi\approx 1/\Gamma$. Other values that we could underestimate are
%total jet power $P_{\rm jet}$ and the relative number of emitting particles $\xi$. Indeed,
%it can be seen that
%\begin{equation}
%\sigma_{\rm M}\propto\frac{\delta}{\left(\sin\varphi\,\xi\right)^{1/3}}P_{\rm jet}^{1/3}.
%\end{equation}
%Here we have used the relationship (\ref{chi}) between half-opening angle and $\varphi$. If
%we assume that $\sin\varphi<1/\Gamma$, and thus $\delta>\beta_{\rm app}$, we conclude that
%Michel magnetization that we have obtained is a lower limit for actual magnetization.

For the highly magnetized regime we come to a contradiction. Indeed, taking, for example,
\nw{a source 0215$+$015, which has Michel magnetization parameter
$\sigma_{0215+015}=\sigma_{\rm M,\,med}$, we obtain from (\ref{Fin3})
for a highly magnetized regime the following value:
\begin{equation}
\sigma_{\rm M,\,mag}=\frac{3.6\cdot 10^{5}}{\Gamma^{3/2}}.
\end{equation}
In highly magnetized regime the scaling (\ref{gamma}) holds, and for
$\Gamma=r_{\perp}/R_{\rm L}\approx10^4-10^5$ we
come to $\sigma_{\rm M,\,mag}\approx 10^{-2}\div 10^{-1}$.
This is in contradiction with our assumption for a magnetized regime with initial magnetization
$\sigma_{\rm M}>10^5$.}

We see that the magnetization parameter $\lambda$ obtained from the observed core-shift has
the order of magnitude $10^{12}\div 10^{14}$ which agrees with the two-photon conversion model
of plasma production in a black hole magnetosphere~\citep{BZ-77, Mosetal-11}. Thus, we obtain
the key physical parameters of the jets being $\sigma_{\rm M} \sim 10$ and $\lambda \sim 10^{13}$.
As a result, knowing these parameters and using rather simple one-dimensional MHD aproach, we
can determine the internal structure of jets.

\begin{figure}
\centering
\includegraphics[scale=0.95]{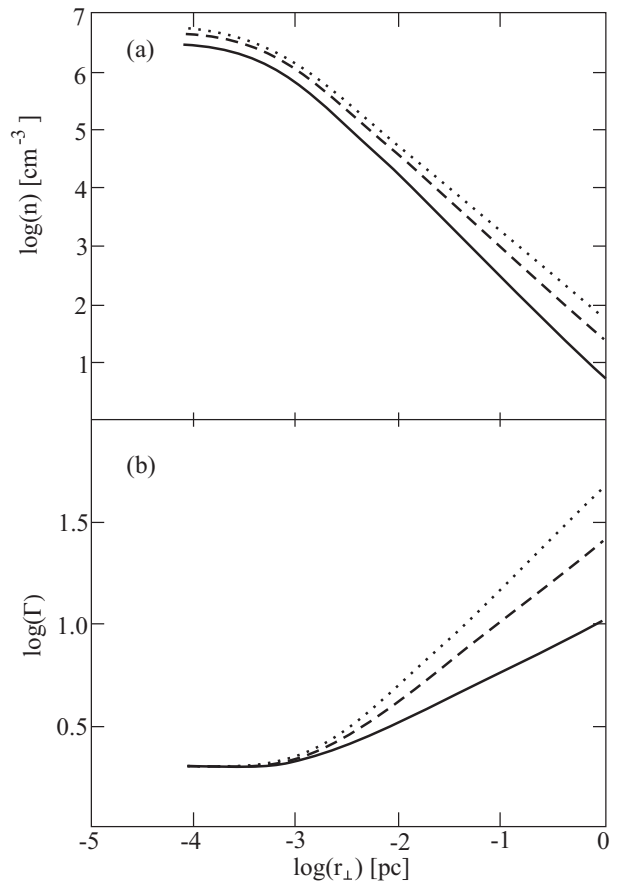}
\caption{
Transversal profile of the number density $n_{\rm e}$ (a) and Lorentz factor $\Gamma$ (b)
in logarithmical scale for $\lambda=10^{13}$, jet radius $R_{\rm jet}=1$ pc and three
different values of $\sigma$: 5 (solid line), 15 (dashed line), and 30 (dotted line).
\label{f:1}
}
\end{figure}

\section{On the internal structure of jets}
\label{s:struct}

As was shown by \citet{BM-00,BN-09,Lyu09}, for well-collimated jets the one-dimensional 
cylindrical MHD approximation (when the problem is reduced to the system of two ordinary 
differential equations, \nw{see above mentioned papers for more detail) allows us to 
reproduce main results obtained later by two-dimensional numerical simulation~\citep{Komissarov-06,
Tchekhovskoy-09, Porth-11, McKinney-12}. In particular, both analytical and numerical
consideration predict the existence of a dense core in the centre of a jet for low enough 
ambient pressure $P_{\rm ext}$. Thus, knowing main parameters obtained above we can determine 
the transverse structure of jets using rather simple 1D analytical approximation. The only
parameters we need are the Michel magnetization $\sigma_{\rm M}$ and the transverse dimension 
of a jet $R_{\rm jet}$ (or, the ambient pressure $P_{\rm ext}$). In particular, transversal 
profiles of the Lorentz factor $\Gamma$, number density $n_{\rm e}$, and magnetic field $B$ 
can be well reproduced. In this section we apply this approach to clarify the real structure
of relativistic jets.}

\begin{figure}
\centering
\includegraphics[scale=0.95]{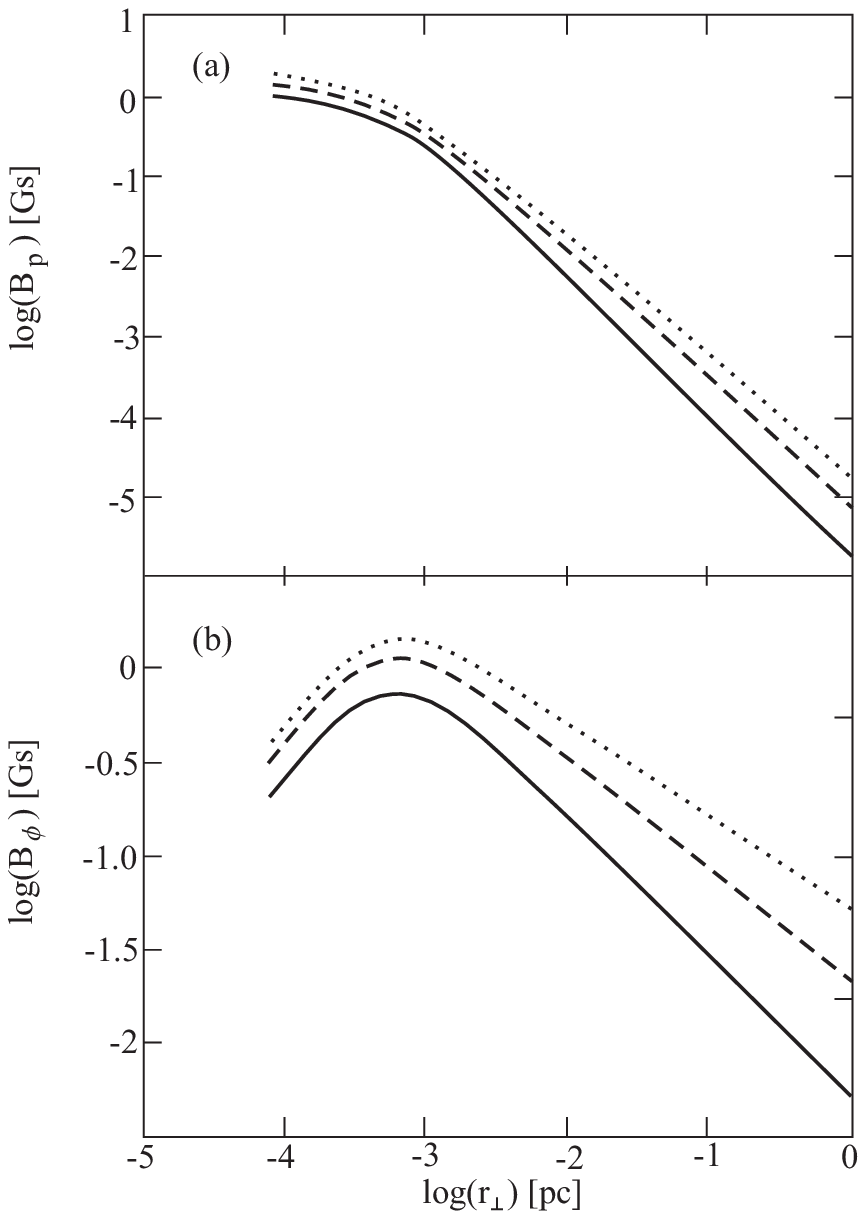}
\caption{
Transversal profile of poloidal (a) and toroidal (b) components of magnetic field
in logarithmical scale for the same parameters and line types as in Fig.~\ref{f:1}.
\label{f:2}
}
\end{figure}

In Fig.~\ref{f:1} we present logarithmic plots of Lorentz factor and number density 
across the jet for $\lambda=10^{13}$, jet radius $R_{\rm jet}=1$ pc and $\sigma=5$, 
15, and 30. Fig.~\ref{f:2} shows logarithmic plots of poloidal and toroidal components 
of magnetic field across the jet with the same parameters as in Fig.~\ref{f:1}.
\nw{As we see, these results point to the existence of more dense central core in the
centre of a jet. Indeed, for our parameters} the number density in the center
of a jet is greater by a factor of a thousand than at the edge. However the
Lorentz factor in the central core is small (see Fig.~\ref{f:1}b). Thus, these results
are in qualitative agreement with previous studies.

\begin{figure}
\includegraphics[scale=0.95]{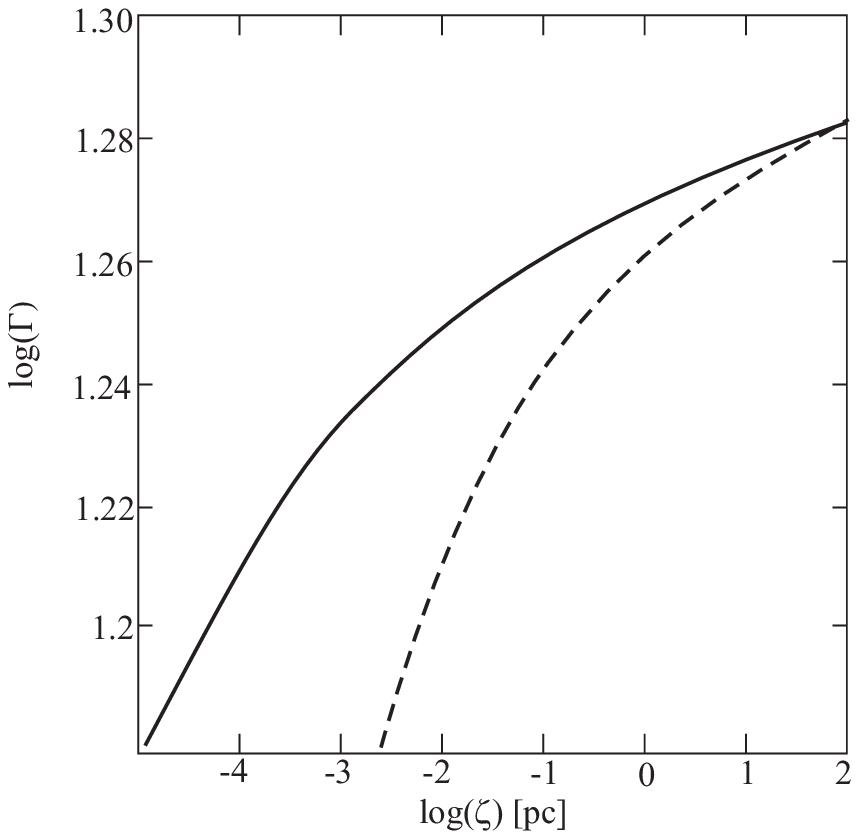}
\caption{
\nw{Dependence of Lorentz factor on coordinate along the jet in assumption of parabolic $\z\propto r_{\perp}^3$ (solid line) 
and $\z\propto r_{\perp}^2$ (dashed line) form of the jet.}
\label{f:9}
}
\end{figure}

Knowing how the Lorentz factor on the edge of jet depends on its radius
and making a simple assumption about the form of the jet, we can calculate
the dependence of the Lorentz factor on the coordinate along the jet. The
result is presented in Fig.~\ref{f:9} for the cases of 
parabolic $\z\propto r_{\perp}^2$ and $\z\propto r_{\perp}^3$ form of the jet. Here
$\z$ is the distance along the axis.
We also assume that the jet has a radius of about
10~pc at the distance 100 pc in both cases, which corresponds to a half-opening angle
of the jet $\theta_{\rm jet}\approx 0.1$. According to Fig.~\ref{f:9}, particle
acceleration in the frame of the AGN host galaxy on the scales 60--100~pc
has values about $\dot{\Gamma}/\Gamma=10^{-3}$ per year with very little dependence of this 
value on the particular form of a jet boundary. This agrees nicely with results of 
the VLBI acceleration study in AGN jets by \cite{Homan-09, Homan-15}.

\section{Studying the causal connectivity of the cylindrical jet model}
\label{s:connect}

The calculated multiplication parameter $\lambda$ with Michel magnetization $\sigma_{\rm M}$ as
well as the observed half-opening angle of a jet $\chi$ allow us to test causal connectivity
across a jet for the cylindrical model. Every spacial point of a super-magnetosonic outflow has
its own ``Mach cone'' of causal influence. In case of a uniform flow the cone originating at the
given point with its surface formed by the characteristics of a flow is a domain, where any signal
from the point is known. For a non-uniform flow the cone becomes some vortex-like shape, depending
on the flow property, but sustaining the property of a causal domain for a given point.

In a jet, if the characteristic inlet from any point of a set of boundary
points reaches the jet axis, we
%, after~\citet{Tchekhovskoy-09}, 
say that the axis is causally connected with the boundary. On the contrary, if there is
a characteristic that does not reach the axis, we have a causally disconnected
flow. In the latter case, a question arises about the self-consistency of
an MHD solution of the flow, since the inner parts of such a flow do not have
any information about the properties of the confining medium. The examples of
importance of causal connectivity in a flow and its connection with the
effective plasma acceleration has been pointed out by \citet{Kom-09} and
\citet{Tchekhovskoy-09}.

In the case of the cylindrical jet model the question of casuality is
even more severe. For a cylindrical model we take into account
the force balance across a jet only, so the trans-field equation
governing the flow becomes one-dimensional. For every initial
condition at the axis its solution gives the flow profile and the position
of a boundary, defined so as to contain the
whole magnetic flux. Any physical value at the boundary such as, for
example, the pressure, may be calculated from this solution. Or,
we in fact reverse the problem, and for a given outer pressure at
the boundary we find the initial conditions at the jet axis.
Thus, we use the dependence of the jet properties at the axis from
the conditions at the boundary. In this case, the boundary and
the axis must be causally connected. In other words, for a strictly
cylindrical flow the conditions at the boundary at the distance
$\z_{0}$ from the jet origin must be ``known'' to the point at the
axis at the same $\z_{0}$.

In the cylindrical model the dynamics of a flow along the jet is achieved by ``piling up''
the described above cross cuts so as to either make the needed boundary form, or to model
the variable outer pressure. In this case the jet boundaries should be constrained by the
``Mach cone'' following causal connection for the model to be self-consistent.Thus, we come
to the following criteria: we may assert that we can neglect the jet-long derivatives in a
trans-field equation if any characteristic, outlet from a boundary at $\z_0$, not only reaches
the axis, but does it at $\z:\, |\z-\z_{0}|\ll \z_{0}$.

\begin{figure}
\centering
\includegraphics[scale=0.95]{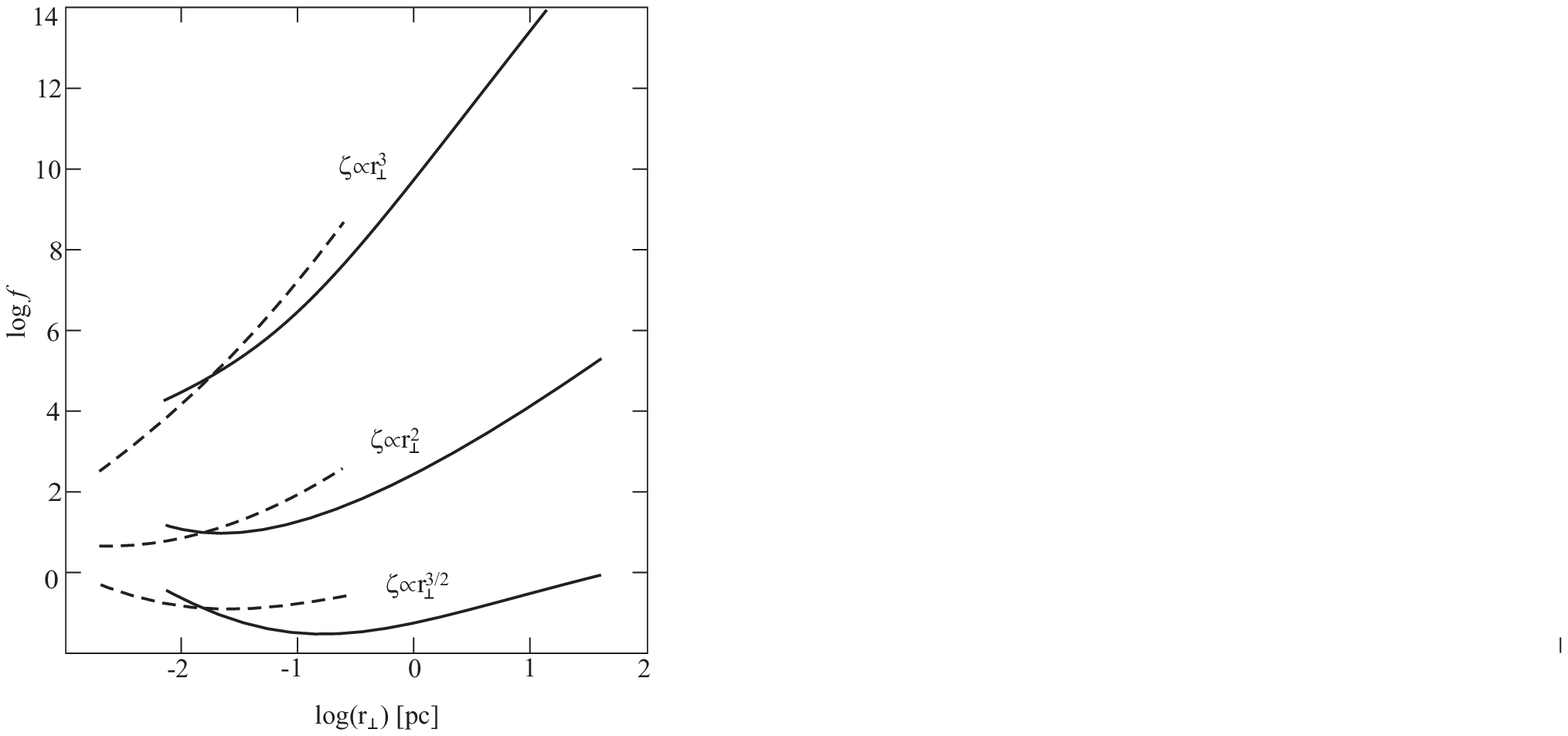}
\caption{The causality function $f$ for different magnetic
surface shapes for $\Gamma>\Gamma_\mathrm{max}/2$, i.e. further the equipartition. 
Solid lines are shown for $\sigma_\mathrm{M}=100$ while dotted --- for $\sigma_\mathrm{M}=10$.
The upper curves correspond to $\z\propto
r_{\perp}^{3}$, the curves in the middle --- to $\z\propto
r_{\perp}^{2}$, and the lower curves --- to $\z\propto r_{\perp}^{3/2}$.
\label{f:f}
}
\end{figure}

For an axisymmetric flow the condition of a causal connectivity across the flow may be
written~\citep{Tchekhovskoy-09} in the simplest case as
\begin{equation}
\theta_{\rm F}>\theta_{\rm j},\label{caus}
\end{equation}
where $\theta_{\rm F}$ is a half-opening angle of a fast Mach cone at the boundary. This
condition means that the characteristic from the jet boundary, locally having its half-opening
angle $\theta_{\rm F}$ with regard to the local poloidal flow velocity, reaches the axis. For
an ultra-relativistic flow, $\theta_{\rm F}$ may be defined as \citep{Tchekhovskoy-09}
\begin{equation}
\sin^2\theta_{\rm F} = \frac{1}{M^2_{\rm F}}=\frac{\Gamma_{\rm
max}-\Gamma}{\Gamma^{3}}.\label{thetaf}
\end{equation}

In the cylindrical approach, we can check the causal connection
across the jet both by applying condition (\ref{caus}), and
by tracking the net of characteristics, outlet from the boundary.
This can be done for a different jet boundary shapes. Let us
introduce the causality function
\begin{equation}
f=\frac{\Gamma_{\rm max}-\Gamma}{\Gamma^{3}}\cdot\frac{1}{\sin^{2}\theta_{\rm j}}.
\end{equation}
It follows from (\ref{caus}) and (\ref{thetaf}) that for $f>1$
causal connectivity holds, and for $f<1$ it does not. If a jet
boundary form is given by a function $\z=\z(r_{\perp})$, where
$r_{\perp}$ is an axial radius, the half-opening jet angle is
defined by
\begin{equation}
\sin\theta_{\rm j} = \frac{\partial r_{\perp}}{\partial \z}
\left[1+\left(\frac{\partial r_{\perp}}{\partial
\z}\right)^{2}\right]^{-1/2}.
\end{equation}
Fig.~\ref{f:f} shows the causality function for a paraboloidal
flow \citep[see][]{BN-09} $\z\propto r_{\perp}^{2}$, for a jet with
a boundary shaped as $\z\propto r_{\perp}^{3}$, and $\z\propto
r_{\perp}^{3/2}$.
For the latter flow shape $f<1$ for every
distance. Thus, the first two outflows are causally connected,
and the last one may be causally disconnected.
Thus, the first two outflows are causally connected, and the last one may be causally disconnected.

\begin{figure*}
\centering
\includegraphics[scale=0.90]{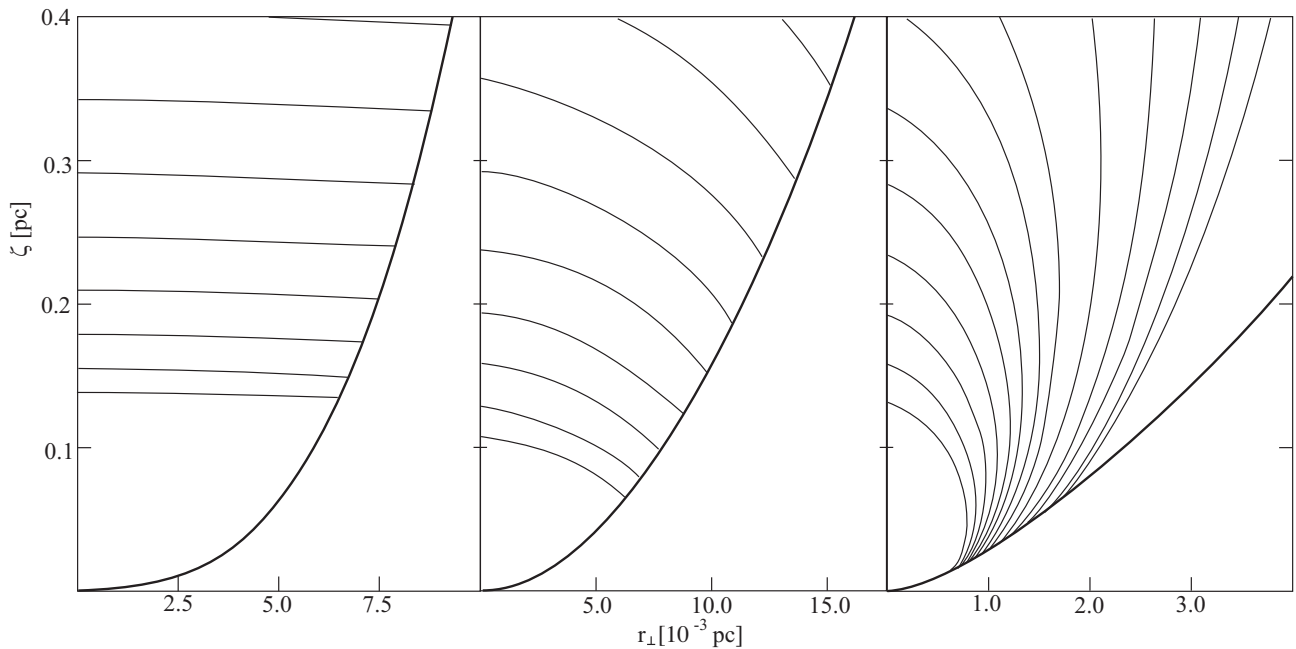}
\caption{The net of inbound characteristics for a parabolic form of a jet boundary
$\z\propto r_{\perp}^{3}$ (left panel),
$\z\propto r_{\perp}^{2}$ (center panel),
$\z\propto r_{\perp}^{3/2}$ (right panel).
\label{f:char-x2}
}
\end{figure*}

The cylindrical approach allows us to investigate the set of
characteristics to check the causality of a flow. Let us discuss
the paraboloidal flow first. We can calculate the Mach
half-opening angle at each point, starting from the boundary, and
so tracking the exact characteristics. This half-opening angle is
defined with regard to the flow velocity direction. Although in
the cylindrical approach all the velocities have only the
$\z$-component, we may introduce the $r_{\perp}$-component by
taking into account the given form of each magnetic surface. The
latter is defined by the function
\begin{equation}
X=\sqrt{r_{\perp}^{2}+\z^{2}}-\z={\rm const},\label{parabola}
\end{equation}
and for inner parts of a flow $\Psi=\Psi_{0}X$. Thus, we define
the angle $\theta_{\ell}$ of a field-line tangent to a vertical
direction as ${\rm tg}\theta_{\ell} =
\left|B_{r_{\perp}}\right|/\left|B_{\z }\right| = \partial
r_{\perp}/\partial \z$, and at the given point of a flow we outlet
the fast characteristic with regard to thus defined flow
direction. We present in Fig.~\ref{f:char-x2} (center panel) the net of
characteristics for a paraboloidal flow depicted as described
above.
The characteristics are calculated starting from the jet boundary towards the axis. They are parameterized by the square of fast magnetosonic Mach number $M_{\mathrm F}^2$ at the axis at the same $\z$, where the characteristic starts. This is done uniformly regarding the Mach number, thus the characteristics are plotted at different distances from each other. 
It can be seen that all the characteristics for parabolic jet boundary reach the axis
at $\z$ not much greater than $\z_{0}$. The same result holds for a
flow $\z\propto r_{\perp}^{3}$ either (see Fig.~\ref{f:char-x2}, left panel). On the contrary, each characteristic, depicted in
Fig.~\ref{f:char-x2} (right panel) for a flow $\z\propto r_{\perp}^{3/2}$,
reaches the axis at a distance along a flow much greater than the
$\z$-coordinate of its origin. This suggests that the cylindrical
approach is definitely not valid in this case.

\section{Discussion}
\label{s:Disc}

We show that the multiplicity parameter $\lambda$, which is the ratio of number density
$n_{\rm e}$ of outflowing plasma to Goldreich-Julian number density $n_{\rm GJ}$, can be
obtained from the direct observations of core shift, apparent opening angle and radio
power of a jet. The formula (\ref{Fin1}) uses the following assumptions, taken from the
theoretical model: (i) the acceleration process of plasma effectivily stops
(saturates) when there is an equipartion regime, i.e. the Poynting flux is equal to
the plasma kinetic energy flux; (ii) we assume the certain power-law scalings for magnetic
field $B(r)$ and number density $n_{\rm e}(r)$ as a functions of distance $r$ \citep{Lob-98}. These scalings
are confirmed by \citet{Sok-11}. We also see that these power-laws are a
good approximation from modelling the internal jet structure in section~\ref{s:struct}.

In contrast with \citet{Lob-98} and \citet{Hir-05} 
we do not assume the equipartition regime of radiating particles
with magnetic field, but the relation between the particles (radiating and non-radiating) kinetic energy and Poynting flux. 
We assume that only the small fraction of particles $\sim 1\%$ radiates 
\citep{SSA-13} and introduce the correlation between particle number density and magnetic field through the flow magnetization 
$\sigma$.  Although for $\sigma\sim 1$ both approaches give effectively the same relation between particle number density and magnetic field, for highly magnetized regime $\sigma \gg 1$ our approach yields the different result.  
Probing both the equipartion regime $\sigma\sim 1$ and highly magnetized regime $\sigma\gg 1$ at parsec scales we conclude that the latter does not hold.

Using the obtained Michel magnetization parameter $\sigma_{\rm M}$, one can easily explain
the observationally derived values \mbox{$2\Gamma\chi \approx 0.1$} \citep{CBetal-13, Zametal-14}, where
$\Gamma$ is Lorentz factor of bulk plasma motion and $\chi$ is a jet half-openong angle. Indeed,
as was found by \cite{Tchekhovskoy-08,Beskin-10}, $2\Gamma\chi \approx 1$ in the whole
domain where $\Gamma \ll \sigma$, independent of the collimation geometry. This implies that
$2\Gamma\chi\approx 1$ up to the distance from the origin whence the transverse
dimension of a jet $R_\mathrm{jet}/R_\mathrm{L} = \sigma_{\rm M}$. At larger distances $\Gamma$
remains practically constant, but for a parabolic geometry the opening angle decreases with the
distance $\zeta$ as $\zeta^{-1/2}\approx r^{-1/2}$. As a result, one can write down
\begin{equation}
2\Gamma\chi\sim \sqrt{\frac{\sigma_{\rm M} R_\mathrm{L}}{R_\mathrm{jet}}}\sim 0.1.
\end{equation}
This result is in agreement with the criteria of casual connectivity across a jet. Indeed,
for an outflow with an equipartition between the Poynting and particle energy flux, we can
write down
\begin{equation}
2\Gamma\chi=\frac{1}{\sqrt{f}}<1
\end{equation}
for a boundary causally connected with an axis.
In Section~\ref{s:connect} we have shown that for flows collimated better than a
parabola, casuality connectivity across the jet holds further, i.e. for $\Gamma>\Gamma_{\rm max}/2$.

\section{Summary}
\label{s:Conc}

The analysis of the frequency dependence of the observed shift of the core of relativistic AGN 
jets allows us to determine physical parameters of the jets such as the plasma number density 
and the magnetic field inside the flow. We have estimated the multiplicity parameter $\lambda$ 
to be of the order $10^{12}$--$10^{15}$. It is consistent with the Blandford-Znajek 
model~\citep{BZ-77} of the electron-positron generation in the magnetosphere of the black hole 
\citep[see][as well]{Mosetal-11}. These values are in agreement with the
particle number density $n_\mathrm{e}$ which was found independently by~\citet{Lob-98}.

As the transverse jet structure depends strongly on the flow regime, whether it is in equipartition 
or magnetically dominated, it is imporatant to know the relation between the observed and maximum 
Lorentz factor. The Michel magnetization parameter $\sigma_\mathrm{M}$ is equal to the maximum 
Lorentz factor of plasma bulk motion.  Typical derived values of $\sigma_\mathrm{M}\lesssim 30$, 
in agreement with the Lorentz factor estimated from VLBI jet kinematics~\citep[e.g.,][]{Cohetal-07, 
Listeretal-09, Listeretal-13} and radio variability~\citep{Jor-05,Hovetal-09,Savetal-10}. This 
implies that a flow is in the saturation regime. Since for strongly collimated flow the condition 
of causial connection is fullfilled (see, e.g.,~\citet{Kom-09, Tchekhovskoy-09}), the internal 
structure of an outflow can be modelled within the cylindrical approach \citep{BM-00, BN-09}. It 
has been shown that the results of the modelling, such as Lorentz factor dependence on the jet 
distance, are in a good agreement with the observations. 
\nw{In particular, the relative growth 
of Lorentz factor $\dot{\Gamma}/\Gamma$ with the distance along the axis is slow for 
the jets in saturation regime, having the magnitude $\sim 10^{-3}$ per year. This result may 
account for the recent masurements of acceleration in AGN jets~\citep{Homan-15}.} 

\nw{
We plan to address the following points in a separate paper:
(i) the role of the inhomogeneity of the magnetic field and particle number density in a core,
(ii) the action of the radiation drag force~\citep{Begelman, BZS-04, Russo},
(iii) the possible influence of mass loading~\citep{ Komissarov-94, SternPoutanen-06, 
DerAharKochKoch-03} on the jet magnetization and dynamics.
}

\section*{Acknowledgments}

We would like to acknowledge E.~Clausen-Brown, D.~Gabuzda, M.~Sikora, \nw{A.~Lobanov, T.~Savolainen, M.~Barkov,
and the anonymous referee} for useful comments.
We thank the anonymous referee for suggestions which helped to imporve the paper.
This work was supported in part by the Russian Foundation for Basic Research grant 13-02-12103.
Y.Y.K.\ was also supported in part by the Dynasty Foundation.
This research has made use of data from the MOJAVE database that is maintained by the MOJAVE
team~\citep{Listeretal-09-MOJAVE}, and data accumulated by the CATS database \citep{Verkh-97}.
This research has made use of NASA's Astrophysics Data System.

\begin{table*}
 \begin{minipage}{150mm}
  \caption{Jet parameters and derived multiplication and magnetization parameters.\label{table}}
  \begin{tabular}{crrrrrcrcrr}
  \hline
  Source & z   & $\beta_\mathrm{app}$ & $\chi_\mathrm{app}$ & $S_{0.3}$ & $P_{\rm jet}$     & Reference     & $\Delta r_{\rm core}$ & Epoch                     & $\lambda$   & $\sigma$ \\
         &     & ($c$)                & ($^\circ$)          & (Jy)      & ($10^{45}$~erg/s) & for $S_{0.3}$ & (mas)                 & for $\Delta r_{\rm core}$ & ($10^{13}$) &          \\
     (1) & (2) & (3)                  & (4)                 & (5)       & (6)               & (7)           & (8)                   & (9)                       & (10)        & (11)     \\
\hline
 0003$-$066 & 0.347 & 8.40 & 16.3 & 2.17 & 1.07 & 2 & 0.035 &2006-07-07& 1.21 & 9.69 \\
 0106$+$013 & 2.099 &  24.37 & 23.6 & 2.85 & 10.50 & 6 &0.005 &2006-07-07& 2.02 & 23.67 \\
 0119$+$115 & 0.570 &  18.57 & 15.6 &   2.24 & 1.86 & 6 & 0.347 &2006-06-15& 3.84 & 4.27 \\
 0133$+$476 & 0.859 &  15.36 & 21.7 &   1.63 & 2.54 & 8 &0.131 &2006-08-09& 3.52 & 5.80 \\
 0202$+$149 & 0.405  & 15.89 & 16.4  &   6.25 & 2.39 & 3 & 0.122&2006-09-06& 1.63 & 10.90 \\
 0202$+$319 & 1.466 & 10.15 & 13.4 &   0.76 & 2.99 & 8  & 0.013&2006-08-09& 2.17 & 11.13 \\
 0212$+$735 & 2.367  & 6.55 & 16.4 &   1.54 & 5.17 & 8 & 0.149&2006-07-07& 9.82 & 4.41 \\
 0215$+$015 & 1.715  & 25.06 & 36.7   & 0.88 & 3.90 & 5 & 0.088&2006-04-28& 3.75 & 7.54 \\
            &       &     &       &     &     &     & 0.241&2006-12-01& 7.97 & 3.54 \\
 0234$+$285 & 1.206  & 21.99 & 19.8   & 1.45 & 3.52 & 5 & 0.275&2006-09-06& 5.29 & 4.79 \\ 
 0333$+$321 & 1.259  & 13.07 & 8.0   & 3.68 & 6.72 & 7 & 0.279&2006-07-07& 4.10 & 8.62 \\
 0336$-$019 & 0.852  & 24.45 & 26.8   & 2.49 & 3.26 & 6 & 0.117&2006-08-09& 2.66 & 8.68 \\
 0403$-$132 & 0.571  & 20.80 & 16.4   & 7.62 & 4.45 & 2 & 0.346&2006-05-24& 3.66 & 6.95 \\
 0420$-$014 & 0.916  & 5.74 & 22.7   & 0.87 & 1.84 & 3 & 0.267&2006-10-06& 13.49 & 1.30 \\
 0458$-$020 & 2.286  & 13.57 & 23.1   & 3.54 & 11.20 & 3 & 0.006&2006-11-10& 3.20 & 17.26 \\
 0528$+$134 & 2.070  & 17.34 & 16.1   & 1.02 & 5.85 & 2 & 0.167&2006-10-06& 4.81 & 7.40 \\
 0529$+$075 & 1.254  & 18.03 & 56.4   & 1.75 & 4.54 & 2 & 0.110&2006-08-09& 3.83 & 7.58 \\
 0605$-$085 & 0.870  & 19.19 & 14.0   & 1.43 & 2.39 & 3 & 0.096&2006-11-10& 1.66 & 11.95 \\
 0607$-$157 & 0.323  & 1.918 & 35.1   & 2.31 & 0.96 & 3 & 0.240&2006-09-06& 28.46 & 0.38 \\
 0642$+$449 & 3.396  & 8.53 & 23.4   & 0.70 & 7.68 & 8 & 0.110&2006-10-06& 5.27 & 8.29 \\
 0730$+$504 & 0.720  & 14.07 & 14.8   & 0.71 & 1.20 & 8 & 0.262&2006-05-24& 4.27 & 3.21 \\
 0735$+$178 & 0.450  & 5.04 & 21.0   & 1.81 & 1.23 & 8 & 0.039&2006-04-28& 2.56 & 5.07 \\
 0736$+$017 & 0.189  & 13.79 & 17.9   & 1.79 & 0.42 & 3 & 0.079&2006-06-15& 0.82 & 8.38 \\
 0738$+$313 & 0.631  & 10.72 & 10.5   & 1.26 & 1.48 & 3 & 0.183&2006-09-06& 2.85 & 5.23 \\
 0748$+$126 & 0.889  & 14.58 & 16.2   & 1.45&  2.65 & 2 & 0.098&2006-08-09& 2.41 & 8.69 \\
 0754$+$100 & 0.266  & 14.40 & 13.7   & 0.74 & 0.39 & 2 & 0.266&2006-04-28& 2.06 & 3.32 \\
 0804$+$499 & 1.436  & 1.15 & 35.3   & 0.60 & 2.49 & 8 & 0.094&2006-10-06& 35.88 & 0.61 \\
 0805$-$077 & 1.837  & 41.76 & 18.8   & 2.60 & 9.26 & 2 & 0.207&2006-05-24& 3.12 & 14.09 \\
 0823$+$033 & 0.505  & 12.88 & 13.4   & 0.63 & 0.71 & 5 & 0.141&2006-06-15& 2.12 & 4.71 \\
 0827$+$243 & 0.942  & 19.81 & 14.6   & 0.71 & 1.80 & 2 & 0.150&2006-05-24& 2.52 & 6.92 \\
 0829$+$046 & 0.174  & 10.13 & 18.7   & 0.67 & 0.21 & 2 & 0.109&2006-07-07& 1.28 & 3.82 \\
 0836$+$710 & 2.218  & 21.08 & 12.4   & 5.06 & 17.75 & 2 & 0.186&2006-09-06& 3.81 & 16.43 \\
 0851$+$202 & 0.306  & 15.14 & 28.5   & 1.11 & 0.56 & 3 & 0.028&2006-04-28& 1.08 & 7.69 \\
 0906$+$015 & 1.026  & 22.08 & 17.5   & 1.54 & 3.05 & 3 & 0.168&2006-10-06& 3.04 & 7.56 \\
 0917$+$624 & 1.453  & 12.07 & 15.9   & 1.27 & 4.07 & 8 & 0.112&2006-08-09& 3.95 & 7.14 \\
 0923$+$392 & 0.695  & 2.76 & 10.8   & 3.28 & 3.03 & 5 & 0.042&2006-07-07& 3.23 & 6.69 \\
 0945$+$408 & 1.249  & 20.20 & 14.0   & 2.94 & 6.30 & 2 & 0.083&2006-06-15& 1.80 & 18.92 \\
 1036$+$054 & 0.473  & 5.72 & 6.5   & 0.75 & 0.80 & 2 & 0.195&2006-05-24& 2.77 & 3.80 \\
 1038$+$064 & 1.265  & 10.69 & 6.7   & 1.59 & 4.32 & 2 & 0.106&2006-10-06& 2.02 & 14.01 \\
 1045$-$188 & 0.595  & 10.51 & 8.0   & 2.79 & 2.46 & 2 & 0.156&2006-09-06& 2.01 & 9.45 \\
 1127$-$145 & 1.184  & 14.89 & 16.1   & 5.63 & 8.94 & 6 & 0.096&2006-08-09& 2.73 & 14.77 \\
 1150$+$812 & 1.250  & 10.11 & 15.0   & 1.39 & 3.81 & 8 & 0.087&2006-06-15& 3.31 & 8.03 \\
 1156$+$295 & 0.725  & 24.59 & 16.7   & 4.33 & 3.89 & 8 & 0.162&2006-09-06& 2.15 & 11.40 \\
 1219$+$044 & 0.966  & 0.82 & 13.0   & 1.14 & 2.52 & 4 & 0.133&2006-05-24& 22.11 & 0.94 \\
 1219$+$285 & 0.103  & 9.12 & 13.9   & 1.77 & 0.19 & 3 & 0.182&2006-02-12& 1.11 & 4.04 \\
            &     &     &       &     &     &     & 0.142&2007-04-30& 0.93 & 4.87 \\
            &     &     &       &     &     &     & 0.199&2006-11-10& 1.19 & 3.78 \\
 1222$+$216 & 0.434  & 26.60 & 10.8   & 3.98 & 1.90 & 5 & 0.180&2006-04-28& 1.14 & 14.13 \\
 1226$+$023 & 0.158  & 14.86 & 10   & 63.72 & 3.25 & 3 & 0.020&2006-03-09& 0.31 & 60.77 \\
 1253$-$055 & 0.536  & 20.58 & 14.4   & 16.56 & 6.31 & 3 & 0.048&2006-04-05& 0.75 & 39.84 \\
 1308$+$326 & 0.997  & 27.48 & 18.5   & 1.42 & 2.79 & 8 & 0.143&2006-07-07& 2.35 & 9.33 \\
 1334$-$127 & 0.539  & 16.33 & 12.6   & 1.91 & 1.71 & 2 & 0.237&2006-10-06& 2.61 & 5.98 \\
 1413$+$135 & 0.247  & 1.78 & 8.8   & 2.74 & 0.81 & 2 & 0.230&2006-08-09& 6.02 & 1.64 \\
 1458$+$718 & 0.904  & 6.61 & 4.5   & 19.64 & 13.30 & 8 & 0.081&2006-09-06& 1.46 & 32.22 \\
            &      &     &      &     &     &     & 0.136&2007-03-01& 2.16 & 21.84 \\ %Sokolovsky \\
 1502$+$106 & 1.839  & 17.53 & 37.9   & 1.08 & 4.92 & 3 & 0.052&2006-07-07& 3.59 & 8.92 \\
 1504$-$166 & 0.876  & 3.94 & 18.4   & 1.80 & 2.79 & 3 & 0.148&2006-12-01& 9.56 & 2.05 \\
 1510$-$089 & 0.360  & 28.00 & 15.2   & 2.75 & 1.22 & 3 & 0.122&2006-04-28& 0.93 & 13.47 \\
 1514$-$241 & 0.049  & 6.39 & 7.8   & 2.06 & 0.08 & 3 & 0.188&2006-04-28& 0.56 & 5.15 \\
\hline
\end{tabular}
\end{minipage}
\end{table*}

\begin{table*}
 \centering
 \begin{minipage}{150mm}
  \contcaption{}
  \begin{tabular}{crrrrrcrcrr}
  \hline
  Source & z   & $\beta_\mathrm{app}$ & $\chi_\mathrm{app}$ & $S_{0.3}$ & $P_{\rm jet}$     & Reference     & $\Delta r_{\rm core}$ & Epoch                     & $\lambda$   & $\sigma$ \\
         &     & ($c$)                & ($^\circ$)          & (Jy)      & ($10^{45}$~erg/s) & for $S_{0.3}$ & (mas)                 & for $\Delta r_{\rm core}$ & ($10^{13}$) &          \\
     (1) & (2) & (3)                  & (4)                 & (5)       & (6)               & (7)           & (8)                   & (9)                       & (10)        & (11)     \\
 \hline
 1538$+$149 & 0.606  & 8.74 & 16.1 &  2.82 & 2.36 & 3 & 0.032&2006-06-15& 1.68 & 11.09 \\
 1546$+$027 & 0.414  & 12.08 & 12.9 &  0.70 & 0.61 & 3 & 0.010&2006-08-09& 0.87 & 10.32 \\
 1606$+$106 & 1.232  & 19.09 & 24.0 &  2.67 & 5.30 & 7 & 0.057&2006-07-07& 2.11 & 14.84 \\
 1611$+$343 & 1.400  & 29.15 & 26.9 &  4.20 & 8.44 & 3 & 0.057&2006-06-15& 1.79 & 22.54 \\
 1633$+$382 & 1.813  & 29.22 & 22.6 &  2.51 & 8.28 & 8 & 0.119&2006-09-06& 3.07 & 13.50 \\
 1637$+$574 & 0.751  & 13.59 & 10.7 &  1.32 & 1.88 & 8 & 0.117&2006-05-24& 1.92 & 8.94 \\
 1638$+$398 & 1.666  & 15.85 & 53.8 &  0.64 & 3.11 & 8 & 0.007&2006-08-09& 4.68 & 5.36 \\
 1641$+$399 & 0.593  & 19.27 & 12.9 &  9.93 & 5.13 & 8 & 0.211&2006-06-15& 2.29 & 11.99 \\
 1655$+$077 & 0.621  & 14.77 & 5.5  &  2.36 & 2.33 & 6 & 0.080&2006-11-10& 0.73 & 25.39 \\
            &        &       &      &       &      &   & 0.086&2007-06-01& 0.77 & 24.05 \\ %Sokolovsky
 1726$+$455 & 0.717  & 2.30 & 16.5 &  0.49 & 0.95 & 8 & 0.009&2006-09-06& 5.18 & 2.34 \\
 1730$-$130 & 0.902  & 27.35 & 10.4 &  6.46 & 6.54 & 3 & 0.174&2006-07-07& 1.67 & 19.72 \\
 1749$+$096 & 0.322  & 7.90 & 16.8 &  1.20 & 0.61 & 6 & 0.061&2006-06-15& 1.43 & 6.15 \\
 1751$+$288 & 1.118  & 3.87 & 12.1 &  0.40 & 1.55 & 2 & 0.007&2006-10-06& 3.60 & 4.62 \\
 1758$+$388 & 2.092  & 2.21 & 17.9 &  0.18 & 1.82 & 8 & 0.079&2006-11-10& 13.98 & 1.42 \\
 1803$+$784 & 0.680  & 10.79 & 18.4 &  1.92 & 2.23 & 8 & 0.029&2006-09-06& 1.71 & 10.80 \\
            &        &      &      &       &      &   & 0.061 &2007-05-03& 1.98 & 9.31 \\ %Sokolovsky \\
 1823$+$568 & 0.664  & 26.17 & 6.8 &  2.63 & 2.52 & 8 & 0.052&2006-07-07& 0.42 & 46.21 \\
 1828$+$487 & 0.692  & 13.07 & 7.1 &  47.78 & 15.60 & 3 & 0.117&2006-08-09& 1.39 & 35.35 \\
 1849$+$670 & 0.657  & 23.08 & 16.6 &  0.86 & 1.22 & 8 & 0.024&2006-05-24& 0.88 & 15.50 \\
 1908$-$201 & 1.119  & 4.39 & 23.9 &  2.70 & 5.21 & 2 & 0.246&2006-03-09& 18.03 & 1.69 \\
 1928$+$738 & 0.302  & 8.17 & 9.8 &  4.81 & 1.40 & 8 & 0.147&2006-04-28& 1.72 & 7.66 \\
 1936$-$155 & 1.657  & 5.34 & 35.2 &  0.67 & 3.45 & 2 & 0.215&2006-07-07& 22.92 & 1.15 \\
 2008$-$159 & 1.180  & 4.85 & 9.7 &  0.73 & 2.41 & 2 & 0.008&2006-11-10& 2.65 & 7.89 \\
 2022$-$077 & 1.388  & 23.23 & 19.6 &  2.63 & 6.67 & 2 & 0.006&2006-04-05& 1.51 & 23.67 \\
 2121$+$053 & 1.941  & 11.66 & 34.0 &  0.63 & 3.99 & 2 & 0.152&2006-06-15& 10.29 & 2.83 \\
 2128$-$123 & 0.501  & 5.99 & 5.0 &  1.47 & 1.23 & 3 & 0.223&2006-10-06& 2.52 & 5.20 \\
 2131$-$021 & 1.284  & 19.96 & 18.4 &  2.66 & 6.11 & 6 & 0.089&2006-08-09& 2.39 & 14.14 \\
 2134$+$004 & 1.932  & 5.04 & 15.2 &  0.99 & 4.85 & 6 & 0.188&2006-07-07& 12.35 & 2.60 \\
 2136$+$141 & 2.427  & 4.15 & 32.5 &  0.94 & 6.16 & 6 & 0.008&2006-09-06& 10.28 & 3.64 \\
 2145$+$067 & 0.999  & 2.83 & 23.2 &  3.76 & 5.18 & 3 & 0.008&2006-10-06& 6.97 & 4.31 \\
 2155$-$152 & 0.672  & 18.12 & 17.6 &  2.41 & 2.43 & 3 & 0.405&2006-12-01& 5.34 & 3.60 \\
 2200$+$420 & 0.069  & 9.95 & 26.2 &  1.82 & 0.12 & 8 & 0.032&2006-04-05& 0.47 & 7.30 \\
 2201$+$171 & 1.076  & 17.66 & 13.6 &  1.00 & 2.63 & 2 & 0.380&2006-05-24& 5.64 & 3.82 \\
 2201$+$315 & 0.295  & 8.27 & 12.8 &  1.82 & 0.88 & 3 & 0.347&2006-10-06& 3.90 & 2.67 \\
            &        &      &      &       &      &   & 0.192&2007-04-30& 2.50 & 3.80 \\
 2209$+$236 & 1.125  & 2.29 & 14.2 &  0.39 & 1.51 & 2 & 0.038&2006-12-01& 6.03 & 2.73 \\
 2216$-$038 & 0.901  & 6.73 & 15.6 &  2.25 & 3.57 & 6 & 0.011&2006-08-09& 2.55 & 9.57 \\
 2223$-$052 & 1.404  & 20.34 & 11.7 &  13.59 & 18.00 & 3 & 0.199&2006-10-06& 3.21 & 18.33 \\
 2227$-$088 & 1.560  & 2.00 & 15.8 &  1.41 & 5.14 & 2 & 0.186&2006-07-07& 22.85 & 1.40 \\
 2230$+$114 & 1.037  & 8.62 & 13.3 &  8.51 & 9.25 & 3 & 0.278&2006-02-12& 7.36 & 5.45 \\
 2243$-$123 & 0.632  & 5.24 & 14.8 &  1.45 & 1.71 & 1 & 0.161&2006-09-06& 5.73 & 2.79 \\
 2251$+$158 & 0.859  & 13.77 & 40.9 &  12.47 & 9.39 & 3 & 0.124&2006-06-15& 8.31 & 4.72 \\
 2345$-$167 & 0.576  & 11.47 & 15.8 &  2.81 & 2.21 & 3 & 0.167&2006-11-10& 3.24 & 5.54 \\
 2351$+$456 & 1.986  & 21.56 & 20.1 &  2.23 & 8.54 & 8 & 0.196&2006-05-24& 5.35 & 7.99 \\
\hline
\end{tabular}
\textit{Notes.} 
Columns are as follows: (1) source name (B1950); (2) redshift $z$ as collected by \citep{Listeretal-13}; (3) apparent velocity measured by \citep{Listeretal-13}; (4) apparent opening angle measured by \citep{Pushetal-09}; (5) flux density at the 92~cm band; (6) derived total jet power; (7) 92~cm flux density reference: 1 --- \citet{WISH}, 2 --- \citet{TXS}, 3 --- \citet{SRCGh}, 4 --- \citet{87GBM}, 5 --- \citet{Ku79r}, 6 --- \citet{Ku81r}, 7 --- \citet{LFVAR}, 8 --- \citet{WENSS}; (8) core shift for frequencies $8.1-15.3$~GHz, measured in mas \citep{Sok-11, Pushetal-12}; (9) an epoch of the core shift measurements by \citet{Pushetal-12} for the year 2006 and by \citet{Sok-11} for the year 2007; (10) derived multiplicity parameter; (11) derived Michel magnetization parameter.
\end{minipage}
\end{table*}

\newpage
\appendix

\section[]{Magnetization parameter}
\label{a:mp}

The standart Grad-Shafranov approach for MHD flows uses the following energy integral $E$, conserved at the magnetic flux surface $\Psi$:
\begin{equation}
E(\Psi)=\frac{c\Omega_{\rm F}(\Psi)I}{2\pi}+m\eta(\Psi)c^2\Gamma.
\end{equation}
Here magnetic field in spherical coordinates $\{r;\,\theta;\,\varphi\}$ is defined by
\begin{equation}
\vec{B}=\frac{\vec{\nabla}\Psi\times\vec{e}_{\varphi}}{2\pi r\sin\theta}-\frac{2I}{cr\sin\theta}\vec{e}_{\varphi},
\end{equation}
electric field
\begin{equation}
\vec{E}=-\frac{\Omega_{\rm F}}{2\pi c}\vec{\nabla}\Psi,
\end{equation}
$\Omega_{\rm F}$ is a rotational velocity and a function of magnetic flux $\Psi$, $I$ is a current, and
\begin{equation}
\eta(\Psi)=\frac{n_{\rm e,*}|\vec{u}_{\rm p}|}{|\vec{B}_{\rm p}|}=\frac{n_{\rm e}|\vec{v}_{\rm p}|}{|\vec{B}_{\rm p}|}.
\end{equation}
Here we assume the flow to be cold, so the particle enthalpy is simply its rest mass.
The Poynting vector
\begin{equation}
\vec{S}=\frac{c}{4\pi}\vec{E}\times\vec{B}=\frac{c}{2\pi}\Omega_{\rm F}I\vec{B}_{\rm p}.
\end{equation}
The particle kinetic energy flux
\begin{equation}
\vec{K}=\left(\Gamma mc^2\right)\left(n_{\rm
e}\vec{v}_p\right)=\Gamma mc^2\eta(\Psi)\vec{B}_{\rm p}.
\end{equation}
Thus, we can introduce the magnetization parameter, variable along the flow, as
\begin{equation}
\sigma=\frac{|\vec{S}|}{|\vec{K}|}=\frac{1}{\Gamma}\frac{1}{2\pi c}\frac{\Omega_{\rm F}I}{m\eta}.
\end{equation}
On the other hand, there is Michel's magnetization parameter $\sigma_{\rm M}$, which has a meaning of magnetization at the
base of an outflow, is constant, and is defined by
\begin{equation}
\sigma_{\rm M}=\frac{E}{mc^2\eta}=\frac{c\Omega_{\rm F}I}{2\pi m\eta c^2}+\Gamma=\Gamma\left(\sigma+1\right).\label{sigma}
\end{equation}

\section[]{Multiplicity parameter}
\label{a:lambda}

The continuity of relativistic plasma number density flux through two cuts, perpendicular to bulk flow, can be written as
\begin{equation}
\lambda n_{\rm GJ} R_{\rm in}^2c=n_1\left(r_1\chi\right)^2c.
\end{equation}
Here we assume the flow velocity equal to $c$.
Using the expression for the Goldreich-Julian concetration
\begin{equation}
n_{\rm GJ}=\frac{\Omega B}{2\pi ce}
\end{equation}
and expression for the full losses due to currents flowing in a magnetosphere
\begin{equation}
P_{\rm jet}=\frac{\left(\Omega B\right)^2 R_{\rm in}^4}{c},
\end{equation}
we get
\begin{equation}
n_1=\frac{\lambda}{2\pi \left(r_1\chi\right)^2}\frac{mc^2}{e^2}\sqrt{\frac{P_{\rm jet}}{P_{\rm A}}}.\label{nWtot_appendix}
\end{equation}

\label{lastpage}

\end{document}